\documentclass[conference]{IEEEtran}
\IEEEoverridecommandlockouts
\usepackage{comment}
\usepackage{mathtools}
\usepackage{bbm}
\usepackage{graphicx}
\usepackage{lipsum}
\usepackage{rotating}
\usepackage{epsfig,graphics,amssymb}
\usepackage{amscd}
\usepackage{amsmath}
\usepackage{enumerate}
\usepackage{lipsum}
\usepackage{mathtools}
\usepackage{theorem}
\usepackage{cite}
\usepackage{stfloats}
\usepackage{epsfig}
\usepackage{verbatim}
\usepackage{color}
\theoremstyle{plain} \theorembodyfont{\itshape}
\theoremheaderfont{\scshape}

\theorembodyfont{\itshape} \theoremheaderfont{\scshape}
\theoremstyle{plain} \theorembodyfont{\itshape}
\theoremheaderfont{\scshape}
\newtheorem{assumption}{Assumption}
\newcommand{\new}{\color{black}}
\title{Dynamical Functional Theory \\ for Compressed Sensing}
\author{
\IEEEauthorblockA{Burak \c{C}akmak \\ Aalborg University \\ 9220 Aalborg, Denmark\\Email: buc@es.aau.dk} \and 
\IEEEauthorblockA{Manfred Opper \\Technical University of Berlin\\Berlin 10587, Germany \\Email: manfred.opper@tu-berlin.de}\and   
\IEEEauthorblockA{Ole Winther \\Technical University of Denmark\\2800 Lyngby, Denmark \\Email: olwi@dtu.dk}  \and 
\IEEEauthorblockA{Bernard H. Fleury \\ Aalborg University \\ 9220 Aalborg, Denmark\\Email: fleury@es.aau.dk} 
\thanks{This work was partially supported by the research project VIRTUOSO funded by Intel Mobile Communications, Keysight, Telenor, Aalborg University, and the Danish National Advanced Technology Foundation.}
}

\begin{document}
\def\mathlette#1#2{{\mathchoice{\mbox{#1$\displaystyle #2$}}%
                               {\mbox{#1$\textstyle #2$}}%
                               {\mbox{#1$\scriptstyle #2$}}%
                               {\mbox{#1$\scriptscriptstyle #2$}}}}
\newcommand{\matr}[1]{\mathlette{\boldmath}{#1}}
\newcommand{\RR}{\mathbb{R}}
\newcommand{\CC}{\mathbb{C}}
\newcommand{\NN}{\mathbb{N}}
\newcommand{\ZZ}{\mathbb{Z}}
\newcommand{\bfl}[1]{{\color{blue}#1}}
\maketitle
\begin{abstract}
We introduce a theoretical approach for designing generalizations of the approximate message passing (AMP) algorithm for compressed sensing which are valid for large observation matrices that are drawn from an invariant random matrix ensemble. By design, the fixed points of the algorithm obey the Thouless-Anderson-Palmer (TAP) equations corresponding to the ensemble. Using a dynamical functional 
approach we are able to derive an effective stochastic process for the marginal statistics of a single 
component of the dynamics. This allows us to design memory terms in the algorithm in such a way that
the resulting fields become Gaussian random variables allowing for an explicit analysis. 
The asymptotic statistics of these fields are consistent with the replica ansatz of the compressed sensing problem. 
\end{abstract}
\section{Introduction}
The canonical problem of compressed sensing is concerned with the linear observation model
\begin{equation}
\matr y=\matr A\matr x+\matr n \label{model}
\end{equation}
where $\matr A$ is the $N\times K$ observation matrix, $\matr x$ is the independent and identically distributed (iid) (unobserved) signal vector, $\matr y$ is the observation vector and $\matr n$ is the iid vector of observation errors whose entries are zero-mean Gaussian with precision $\xi$. Here, all variables are real-valued. The dimensions $N$ and $K$ are both large, possibly arbitrary large with the aspect ratio $\alpha\triangleq N/K$ fixed. Compressed sensing aims to recover the signal $\matr x$ from a compressive sampling system, i.e. $N<K$, by exploiting the sparsity structure of $\matr x$.

Within the above context, the approximate message passing (AMP) algorithm \cite{Donoha} has drawn significant attention in the communities of information theory and signal processing. In particular, the algorithm has the virtue that the trajectories of its dynamics can be described by a simple \emph{one-dimensional} iterative equation -- called state evolution\cite{Bayati}. State evolution is not only useful for performing the signal recovery but also for the analysis of \eqref{model} within information theory \cite{Barbier16}. 

The AMP method is based on the assumption that the entries of observation matrix are iid zero-mean Gaussian\footnote{The Gaussian condition can be relaxed to a sub-Gaussian distribution\cite{bayati15}.} with variance $1/N$. However, without a clear physical and/or mathematical motivation, the iid assumption can be artificial. Indeed, signals are often sparse in the Fourier domain, in which case the observation matrix is typically a $N\times K$ ``corner'' of the $K\times K$ randomly-permuted discrete-cosine transform (DCT) matrix. In this case, the optimality of the AMP algorithm is broken and state evolution becomes invalid. 

It is the goal of this study to develop a theoretical framework that extends the AMP method in such a way that it is valid for general invariant matrix ensembles \cite{Percy}. Specifically, we assume that Gramian $\matr {\tilde J}\triangleq\matr A^\dagger \matr A$ is invariant, i.e. it has the same distribution as $\matr O^\dagger\matr {\tilde J}\matr O$ for any orthogonal matrix $\matr O$ independent of $\matr{\tilde J}$.\footnote{Though a Gramian of the corner of randomly-permuted DCT matrix is not invariant, it asymptotically behaves as an invariant matrix, see e.g.\cite{Enzo,Greg}.} These ensembles are now favoured in performance analyses of compressed sensing \cite{Tulino13,Mikko,Ali16}. 

The AMP algorithm can be viewed as a system of dynamical equations that attempt to solve the so-called Thouless-Anderson-Palmer (TAP) (fixed-point) equations \cite{Mezard,Adatap,OW5} of the observation model \eqref{model} under the iid assumption about the entries of $\matr A$, see \cite{Kab14,Samp,Burak16}. Although the AMP method was originally derived under the assumption that the observation matrix is sparse, due to the iid property of the matrix elements, the method also leads to a correct algorithm when said matrix is dense. Since it is not clear how to deal with sparse observation matrices for more general random matrix ensembles, we develop in this study a method for describing the dynamics of iterative algorithms involving dense observation matrices as a starting point. Specifically, we first introduce general dynamical equations of a candidate algorithm that attempts to solve the TAP equations for general invariant matrix ensembles. Then, we make use of dynamical functional theory (DFT)\cite{Martin,Eisfeller} to statistically characterize the trajectory generated by these equations in the large-system limit (more specifically, in the thermodynamic limit). This approach yields a method which allows for both designing and analyzing algorithms involving general invariant matrix ensembles.  

This paper is organized as follows: In Section~2 we present the elementary tools from random matrix theory that we use in this contribution and summarize the TAP equations of \eqref{model} for general invariant ensembles. In Section~3 we present the results of DFT. Section~4 presents the ``single-step memory'' algorithm. Section~5 is devoted to the derivation of the statistical characterization of the trajectory of the single-step memory algorithm. Section~6 gives a summary and outlook. 

\subsection{Related works}
We refer to \cite{Mimura} which makes use of DFT to analyze an AMP-type iterative detection algorithm for code-division--multiple-access systems with the classical iid assumption about the entries of the observation matrix.

From the technical point of view, this paper can be viewed as an extension of \cite{Opper16} where DFT is used for solving the TAP equations of the Ising model of spin glass theory. Due to space limitations, we do not present the proofs of our results; they are obtained by following arguments similar to those used in \cite{Opper16}, see also the paragraph below \eqref{gen}.

\subsection{Notations}
The entries of a $K\times N$ matrix $\matr X$ and a $K\times 1$ vector $\matr x$ are denoted {\new as} $X_{nk}$ and $x_k$ for $n\in [1,N]$ and $k\in [1,K]$. We use calligraphic notation, e.g. $\mathcal G$, for $T\times T$ matrices whose entry indices range (for convenience) in {\new $[0,T-1]$}, i.e. starting at $0$. Thus, $\mathcal G(t,\tau)$ is the $(t+1,\tau+1)$ entry of $\mathcal G$.
The transposition is denoted by $(\cdot)^\dagger$.
The Gaussian distribution function with mean $\matr \mu$ and covariance matrix $\matr \Sigma$ is denoted by $\mathcal N(\cdot ; \matr \mu, \matr\Sigma)$. 
For random variables $x$ and $y$, $x\sim y$ implies that $x$ and $y$ are identically distributed. We indicate that $x$ has a probability density function (pdf) $p$, or distribution function ${\rm P}$, by $x\sim p$, or $x\sim {\rm P}$. Moreover, $\langle x\rangle=\langle x\rangle_p= \int x {\rm dP}(x)$.

\section{Preliminaries}
Throughout the paper, we assume the following.
\vspace{-0.2cm}
\begin{assumption}
Gramian $\matr {\tilde J}=\matr {A}^\dagger\matr A$ is invariant, has a uniformly bounded spectral norm and its empirical eigenvalue distribution converges almost surely to a limiting eigenvalue distribution (LED) as $K\to \infty$ (with $\alpha=N/K$ fixed). For convenience, we assume that ${\rm tr}(\matr {\tilde J})/K\to 1$ almost surely as $K\to \infty$.  
\end{assumption}
\vspace{-0.1cm}

We shall exemplify our general results through two specific matrix ensembles: (i) $\matr A$ has iid zero-mean Gaussian entries with variance $1/N$; (ii) $\matr A$ is random row-orthogonal, namely $\matr A= \alpha^{-\frac 1 2}\matr P_{\alpha} \matr O$ where $\matr O$ is a $K\times K$ Haar matrix and $N \times K$ matrix $\matr P_{\alpha}$ removes $K-N$ rows of $\matr U$. Specifically, $[P_{\alpha}]_{ij}=\delta_{ij}$, $\forall i,j$, where $\delta_{ij}$ denotes the Kronecker delta.

For mathematical convenience, our random matrix arguments will be based on the LED of $\matr J\triangleq\xi{\bf I} - \xi\matr {\tilde J}$. 

\subsection{R-transform and its inverse}
Our key tool from random matrix theory is the so-called R-transform \cite{Tulino} {\new of the LED of $\matr J$ that is defined as}
\begin{equation}
{\rm R}(\omega)\triangleq\sum_{n=1}^{\infty}c_n\omega^{n-1}
\end{equation} 
where $c_n$ is the distribution's $n$th free-cumulant. Here, we underline the simple relationship ${\rm R}(\omega)=\xi-\xi{\rm {\tilde R}}(-\xi \omega)$ where ${\rm {\tilde R}}$ is the R-transform of the LED of $\matr {\tilde J}$. In particular, ${\rm {\tilde R}}$ can be conveniently obtained from the inverse (with respect to the composition of functions) of the Stieltjes transform of the LED of $\matr {\tilde J}$, for details we refer to \cite{Tulino}. 

For matrix ensemble (i), ${\rm {\tilde R}}$ is the R-transform of the Mar\u{c}enko-Pastur distribution (see \cite{Tulino}) and thereby we have
\begin{equation}
{\rm R}(\omega)= \frac{\xi^2\omega}{\alpha+\xi \omega}.\label{Riid}
\end{equation}
For matrix ensemble (ii), one can show that (see \cite[Eq. (36)]{Tulino13})
\begin{equation}
{\rm R}(\omega)=\xi-\frac{(\frac{\xi \omega}{\alpha}+1)-\sqrt{(\frac{\xi \omega}{\alpha}+1)^2-4\xi \omega}}{2\omega}.
\end{equation}

By the definition of $\matr J$, we have ${\rm R}(0)=0$. Hence, Lagrange's inversion theorem implies that the inverse ${\rm R}^{-1}$ exists and is analytic in a neighborhood of zero. Thus, we can write 
\begin{equation}
{\rm R}^{-1}(\omega)=\sum_{n=1}^{\infty}a_n\omega^n. \label{inv}
\end{equation}
For matrix ensembles (i) and (ii) we obtain respectively $a_{n}=(\alpha\xi^{-1})\xi^{-n}$ and 
\begin{equation}
\quad a_n=(\alpha \xi^{-1})\xi_1^{-n}-(\alpha \xi^{-1})\xi_2^{-n}\label{memrow}
\end{equation} 
where for short we introduce the vector $\matr\xi\triangleq(\xi,\xi(\alpha-1)/\alpha)$. 

\subsection{TAP equations with general invariant  matrix ensembles}
The TAP method \cite{Mezard,Adatap} -- also known as expectation consistency in machine learning\cite{OW5} -- typically provides highly accurate approximations for probabilistic inference. In our context, the TAP equations refer to an approximate, denoted by $\matr m$, of the minimum-mean-square-error estimate of the signal, i.e.~the mean of the posterior pdf of $\matr x$: $\matr m\approx \langle \matr x \rangle_{p(\matr x\vert \matr y)}$.

For convenience, we introduce the auxiliary pdf
\begin{equation}
q_{\matr \psi;v}(\matr x)\propto p(\matr x) \exp\left(-\frac{v}{2}\Vert\matr x -\matr \psi\Vert^2\right)\label{priord}
\end{equation}
where $p(\matr x)$ is the prior pdf of the iid signal vector $\matr x$. The mean and the normalized total variance of this pdf are denoted as
\begin{align}
\eta_v(\matr \psi)\triangleq \langle \matr x \rangle_{q_{\matr \psi;v}} ~\text{and}~ \chi\triangleq\frac{1}{K}\langle \Vert \matr x-\eta_v(\matr \psi)\Vert^2\rangle_{q_{\matr \psi; v}}.\label{varq}
\end{align}
Then, the TAP equations of \eqref{model} for general invariant matrix ensembles can be expressed as \cite{Adatap,Burak16}
\begin{subequations}
	\label{TAP}
	\begin{align}
	\matr m &= \eta_v(\matr \psi) \quad \text{with} \quad v\triangleq{\xi-{\rm R}(\chi)}  \\
	\matr \psi&=\frac{1}{v}\left(\matr h +\matr J\matr m-{\rm R}(\chi)\matr m\right)\quad \text{with} \quad\matr h\triangleq\xi \matr A^\dagger \matr y.  \label{TAP2}
	\end{align}
\end{subequations}
Note that the dependency on the random matrix ensemble in the TAP equations is via the R-transform of the LED of $\matr J$. 

The TAP equations are known to be consistent with the replica ansatz \cite{Tulino13} -- which is commonly assumed to be exact above the so-called Almedia-Thouless line of the stability \cite{AT}. Specifically, the entries of the static field $\matr \psi$ are asymptotically iid --  as implied by the decoupling principle -- and \cite{Kab14}
\begin{equation}
\psi_k\sim \theta +x \quad \text{with} \quad \theta\sim\mathcal N(0,1/v)~ \text{and}~x\sim p(x) \label{replica}
\end{equation}
for $\theta$ independent of $x$. {\new Here and in the sequel, we substitute $v$ (or $\chi$) with its (non-random) large-system approximation obtained with the replica ansatz, i.e. $\chi=\langle(x-\eta_{v}(\theta +x))^2\rangle_{x,\theta}$}. We next extend the ``decoupling principle'' originally formulated for static solutions to the trajectory of a dynamical algorithm that computes these solutions.
\section{The Results of Dynamical Functional Theory}
We start with the following system of dynamical equations of the discrete time $t=0,1,\ldots,T-1$ as a candidate for solving the TAP equations \eqref{TAP}:
\begin{subequations}
\label{dynamic}
\begin{align}
\matr m(t) &= f_{t}\left(\{\matr \gamma(\tau),\matr m(\tau)\}_{\tau<t}\right) \label{good2}\\
\matr \gamma(t) &= \matr h+ \matr J\matr m(t)\label{good}
\end{align} 
\end{subequations}
where $\{f_t\}_{t<T}$ is an appropriate sequence of non-linear scalar functions. We obtain the statistics of the trajectory of this system from its generating functional 
\begin{align}
Z(\{\matr l(t)\})=\int&\prod_{t=0}^{T-1}{\rm d}\matr m(t){\rm d}\matr \gamma(t)\;e^{
{\rm i}
{\matr \gamma(t)^\dagger \matr l(t)}}\times\nonumber\\&\times \delta(\matr m(t)- f_t\left(\{\matr\gamma(\tau), \matr m(\tau) \}_{\tau<t}\right))\times \nonumber \\&\times\delta(\matr \gamma(t)-\matr h -\matr J\matr m(t)). \label{gen}  
\end{align}
Note that $\matr h$ is a function of the random elements $\matr x$, $\matr n$ and $\matr A$, specifically $\matr h=\xi\matr A^\dagger(\matr A\matr x+\matr n)$. This is what essentially distinguishes the problem from that addressed in \cite{Opper16} and requires the proofs in \cite{Opper16} to be adapted. 

We are interested in the thermodynamical properties of $Z(\{\matr l(t)\})$, i.e. its expectation over $\matr x$, $\matr n$ and $\matr A$ that we denote as
$\langle Z(\{\matr l(t)\})\rangle$. Here, we note that $Z(\{\matr l(t)=\matr 0\})=1$ and therefore direct averaging of  $Z(\{\matr l(t)\})$ is a proper operation. By making use of the R-transform formulation of
the asymptotic Itzykson-Zuber integral \cite{Collins5} and the saddle-point method\cite{Neri} we obtain that $\langle Z(\{\matr l(t)\})\rangle$ factorizes (or decouples) as
\begin{align}
\hspace*{-2ex}
\prod_{k=1}^K\int {\rm dP}(x_k) {\rm d}\mathcal N (\{{\phi_k}(t)\}; \matr 0,\mathcal C_{\phi})\; \prod_{t=0}^{T-1}{\rm d}m_k(t){\rm d}\gamma_k(t)\times \nonumber\\\vspace{0.6cm}\times 
e^{O(\frac 1 K)+
{\rm i}
{\gamma_k(t)l_k(t)}}\times\nonumber \\ \times\delta (m_k(t)- f_t\left(\{\gamma_k(\tau), m_k(\tau) \}_{\tau<t}\right))\times \nonumber \\ \vspace{0.3cm}\times\delta\Big(\gamma_k(t)-\phi_k(t)-\xi x_k -\sum_{\tau= 0}^{t-1} \mathcal {\hat G}(t,\tau)(m_k(\tau)-x_k)\Big)
\label{E-EGF}
\end{align}
where $O(1/K)$ is a constant term that, as indicated, vanishes as $K\to\infty$. The random fields $\{\phi_k(t)\}=\{\phi_k(t)\}_{t<T}$, $k\in [1,K]$ have the ($T\times T$) covariance matrix
\begin{align}
\mathcal C_{\phi}=\xi\matr 1 \matr 1^\dagger&-\sum_{n=1}^{\infty}c_n \sum_{k=0}^{n-1} \mathcal {G}^k\matr 1 \matr 1^\dagger(\mathcal {G}^\dagger)^{n-1-k}\nonumber \\&+\sum_{n=1}^{\infty}c_n \sum_{k=0}^{n-2} \mathcal {G}^k\mathcal {C}(\mathcal {G}^\dagger)^{n-2-k}. \label{cov}
\end{align}
Here, $\matr 1$ is the all-ones $T\times 1$ vector and $\mathcal{G}$ and $\mathcal{C}$ are the $T\times T$ response and correlation matrices respectively with 
\begin{align}
\mathcal G(t,\tau)&=\left<\frac{\partial m_k(t)}{\partial \phi_k(\tau)}\right> \label{response}\\
\mathcal C(t,\tau)&=\left<(m_k(t)-x_k)(m_k(\tau)-x_k)\right>. \label{corr}
\end{align}
The expectations in the above expressions are over random variables $\phi_k$ and $x_k$. The $T\times T$ memory matrix $\mathcal {\hat G}$ in (\ref{E-EGF}) is defined as
\begin{equation}
\mathcal{\hat G}= {\rm R}(\mathcal{G}).\label{mem}
\end{equation}
These results imply that for sufficiently large $K$, we can compute the marginal statistics (i.e. 
the statistics of single components of the vector $\matr m(t)$) in \eqref{dynamic} from
an effective stochastic process with iid components that is given by
\begin{subequations}
\begin{align}
\matr m(t)&= f_{t}\left(\{\matr \gamma(\tau),\matr m(\tau)\}_{\tau<t}\right) \label{mdym}\\
\matr \gamma(t) & =\matr \phi(t)+\xi \matr x+\sum_{\tau=0}^{t-1}\mathcal {\hat G}(t,\tau)(\matr m(\tau)-\matr x) \label{rdym} 
\end{align}	
\end{subequations}
where the entries of $\matr \phi(t)$ are iid zero-mean Gaussian processes with $T\times T$ covariance matrix $\mathcal C_{\phi}$. Unfortunately, due to the occurrence of the memory terms in (\ref{rdym}) the field $\matr \gamma(t)$ becomes non--Gaussian which makes an analysis of the algorithm in general intractable. While there are many ways to overcome this problem, we next limit our attention to a particular design in which almost all memory terms disappear.
\section{The Single-Step Memory Design}
We consider (\ref{mdym}) to be of the form $\matr m(t+1)=\eta_{v(t)}(\matr \psi(t))$. Here, the scalar $v(t)$ is non-random. As long as $v(t)\to v$ is guaranteed we have the freedom to choose the dynamic evolution of $v(t)$ without referring to the results of DFT. We will use the results of DFT to design the random field $\matr \psi(t)$ in such a way that its components $\{\psi_k(t):k\in [1,K]\}$
are iid but ``temporally'' dependent.

In the single-step memory design the dynamical system is built on the basis of the single-step memory condition\cite{Opper16}
\begin{equation}
\mathcal {\hat G}(t,\tau)=0, \; \forall \tau\neq t-1. \label{ssm}
\end{equation}
From \eqref{good} and \eqref{rdym} this implies that  
\begin{align}
\matr \phi (t)&=\matr \gamma(t)-\xi \matr x-\mathcal {\hat G}(t,t-1)(\matr m(t-1)-\matr x) \\
&= \matr h+\matr J\matr m(t)-\mathcal {\hat G}(t,t-1)\matr m(t-1)- \nonumber \\
& -(\xi-{\mathcal {\hat G}(t,t-1)})\matr x
\end{align}
with $\mathcal {\hat G}({0,-1})\triangleq0$. We design the field $\matr \psi(t)$ to be of the form
\begin{align}
\matr \psi (t)&= \sum_{\tau=0}^{t}\mathcal A(t+1,\tau)\left[\matr \phi (\tau)+(\xi-{\mathcal {\hat G}(\tau,\tau-1)})\matr x \right] \label{trans}  \\
 &=\sum_{\tau=0}^{t}\mathcal A(t+1,\tau)[\matr h+\matr J\matr m(\tau)-\mathcal {\hat G}(\tau,\tau-1)\matr m(\tau-1)].\nonumber 
\end{align}
Conditioned on the static random field $\matr x$, the field $\matr \psi(t)$ is Gaussian, which allows for an explicit analysis. Basically, we have to design the non-random terms $\mathcal A(t,\tau)$ in such a way that the construction is consistent with the results of DFT. From \eqref{trans} we obtain the relationship
\begin{align}
\mathcal G(t,\tau)&= \left\langle\frac{\partial \eta_{v(t-1)}(\psi_k(t-1))}{\partial \phi_k(\tau)} \right  \rangle\\ &= {\chi(t)}v(t-1)\mathcal A(t,\tau) \quad \text{with}~\chi(t)\triangleq \mathcal C(t,t).
\end{align}
Furthermore, from \eqref{mem} we have $\mathcal G={\rm R}^{-1}(\mathcal {\hat G})$. Combining with the single-step memory property \eqref{ssm} yields
\begin{equation}
\mathcal G(t,\tau)=a_{t-\tau}\prod_{s=\tau}^{t-1}\mathcal {\hat G}(s+1,s).
\end{equation}
Finally, we need to specify the memory terms $\mathcal {\hat G}(t,t-1)$ in such a way that the resulting algorithm is asymptotically (as $t\to \infty$) consistent with the TAP equations. We choose
\begin{equation}
\mathcal {\hat G}(t,t-1)=\frac{\chi(t)}{\chi(t-1)}{\rm R}(\chi(t-1)). 
\end{equation}
Combining the above steps, the single-step memory algorithm (for $t\geq 0$) can be written in the form
\begin{subequations}
\label{ssma}
\begin{align}
\matr m(t+1)&=\eta_{v(t)}(\matr \psi(t)) \\
\matr \psi (t)&= \frac{Q(t)}{v(t)}\sum_{\tau=0}^{t}a_{t+1-\tau}\matr u(\tau) \\
\matr u(t)&=\frac{\matr h+\matr J\matr m(t)-\mathcal {\hat G}(t,t-1)\matr m(t-1)}{\chi(t)Q(t-1)}
\end{align}	
\end{subequations}
where $Q(t)=Q(t-1){\rm R}(\chi(t))$ and $Q(-1) {\triangleq} 1$. 
\subsection{Asymptotic consistency with the TAP equations} 
Following the arguments \cite{Opper16} one can show that if the single-step memory algorithm \eqref{ssma} converges it solves the TAP fixed-point equations \eqref{TAP} under the so-called weak long-term response assumption \cite{Kurchan}
\begin{equation}
\lim_{t \to \infty}\mathcal{G}(t,\tau)=0\quad \forall~\text{finite}~\tau \label{weaklong}.
\end{equation}
One can also show that \eqref{weaklong} holds for matrix ensembles (i) and (ii). Instead, in the sequel we further simplify the form of the single-step memory algorithm for these ensembles in such a way that the consistency of its fixed points with the TAP equations can be trivially checked. We first consider the matrix ensemble (ii).

\subsection{Example-1: Matrix ensemble~(ii)}
By using the expression of coefficient $a_n$ for matrix ensemble (ii) in \eqref{memrow}, we decompose the field $\matr \psi(t)$ as
\begin{align}
\matr\psi(t)=\underbrace{\frac{Q(t)}{v(t)}\frac{\alpha}{\xi}\sum_{\tau=0}^{t}\frac{\matr u(\tau)}{\xi_1^{t+1-\tau}}}_{\matr \psi_1(t)}-\underbrace{\frac{Q(t)}{v(t)}\frac{\alpha}{\xi}\sum_{\tau=0}^{t}\frac{\matr u(\tau)}{\xi_2^{t+1-\tau}}}_{\matr \psi_2(t)}.\label{field}
\end{align}
It is easy to see that each field admits the recursion
\begin{align}
\matr \psi_{i}(t)
=&\frac{\alpha}{\xi}\frac{{\rm R}(\chi(t))}{\chi(t)v(t)}\frac{1}{\xi_i}(\matr h+\matr J\matr m(t)-\mathcal {\hat G}(t,t-1)\matr m(t-1))\nonumber \\
&+\frac{{\rm R}(\chi(t))v(t-1)}{v(t)}\frac{1}{\xi_i}\matr \psi_{i}(t-1).\label{recur}
\end{align}
For further convenience, we introduce the auxiliary fields
\begin{equation}
 \matr z_i(t)\triangleq \frac{\matr\psi_i(t)}{c(t)}-\matr m(t)\quad \text{with}\quad c(t)\triangleq 
\frac{\alpha}{\xi}\frac{{\rm R}(\chi(t))}{\chi(t)v(t)}.\label{aux}
\end{equation}
Applying this transformation to the recursion \eqref{recur}, the single-step memory algorithm \eqref{ssm} can be written in the form
\begin{align}
\matr m(t+1)&=\eta_{v(t)}\left(c(t)[\matr z_1(t)-\matr z_2(t)]\right)\nonumber \\
\matr z_{i}(t)&=\frac{1}{\xi_i}(\matr h+\matr J\matr m(t)-\xi_i\matr m(t)+\mathcal {\hat G}(t,t-1)\matr z_i(t-1)).\label{ssma2}
\end{align} 
\subsection{Example-2: AMP algorithm}
For matrix ensemble (i), we show that by choosing the dynamic evolution of $v(t)$ as
\begin{equation}
v(t)=\xi-{\rm R}(\chi(t))= \frac{\xi\alpha}{\alpha+\xi\chi(t)},\label{amp1}
\end{equation}
the single-step memory algorithm yields the AMP algorithm \cite{Donoha,Bayati}: We have  $a_{n}=(\alpha\xi^{-1})\xi^{-n}$. From the previous discussion, this implies that $\matr \psi(t)=c(t)(\matr z(t)-\matr m(t))$ where $\matr z(t)$ admits exactly the same form as $\matr z_1(t)$ in \eqref{ssma2} and $c(t)$ is as in \eqref{field}. Moreover, \eqref{amp1} yields that $c(t)=1$. Thereby, the single-step memory algorithm \eqref{ssm} simplifies to
\begin{align}
\matr m(t+1)&=\eta_{v(t)}(\matr z(t)+\matr m(t))\\
\matr z(t)&=\matr A^\dagger(\matr y-\matr A\matr m(t))+ \frac{1}{\xi} \mathcal {\hat G}(t,t-1)\matr z(t-1).
\end{align}
\section{The Field Statistic}
To describe the trajectory of the single-step memory algorithm we solely need an explicit statistical characterization of the random field $\matr \psi(t)$. We note that its entries are iid and are expressed as linear combinations of the corresponding entries of the Gaussian field $\matr \phi(t)$ (see \eqref{trans}): 
\begin{equation}
\underbrace{\sum_{\tau=0}^t\mathcal A(t+1,\tau)\phi (\tau)}_{\theta(t)}+{x}\underbrace{\sum_{\tau=0}^t\mathcal A(t+1,\tau)(\xi-{\mathcal {\hat G}(\tau,\tau-1)})}_{\sigma_x(t)}\nonumber,
\end{equation}
i.e. $\psi_{k}(t)\sim \theta(t)+\sigma_{x}(t)x$, with $\{\phi(t)\}\sim{\mathcal N}(\matr 0,\mathcal C_{\phi})$, see \eqref{cov}. The expression $\sigma_{x}(t)$ can be written in the compact form
\begin{equation}
\sigma_{x}(t)=\frac{1}{v(t)}\left[\zeta(t)\xi-\zeta(t-1){\rm R}(\chi(t))\right]. \label{sigmax}
\end{equation}
Here, $\zeta(-1)\triangleq0$ and for ${t\geq0}$ we have
\begin{equation}
\zeta(t)\triangleq Q(t)\sum_{\tau=0}^t\frac{a_{t+1-\tau}}{\chi (\tau)Q(\tau-1)}.
\end{equation}
Thus, we merely need to derive an explicit expression of the covariance matrix of $\{\theta(t)\}$, say $\mathcal C_\theta$. To this end, we introduce the bi-variate random matrix transformation
\begin{equation}
{\rm B}(\omega,z)\triangleq \left(\frac{1}{{\rm R}^{-1}(\omega)}-\frac{1}{{\rm R}^{-1}(z)}\right)^{-1}(z-\omega) \label{Afunction}.
\end{equation}
Moreover, for power series $f(\omega,z)=\sum_{n,k\geq 0}a_{n k} \omega^nz^k$, ${\rm Co}_{\omega^nz^k} [f(\omega,z)]\triangleq a_{n k}$. With these definitions, we obtain the explicit expression of  $\mathcal C_\theta$ given in \eqref{covmax} at the top of the next page.
\begin{figure*}[!t]
	\normalsize
	\setcounter{equation}{38}
	\begin{equation}
\mathcal C_\theta(t,t')=\frac{\zeta(t')}{v(t')}\sigma_x(t)+ \frac{Q(t)Q(t')}{v(t)v(t')} \sum_{\tau=0}^{t}\sum_{s=0}^{t'}\frac{{\rm Co}_{\omega^{t+1-\tau}z^{t'+1-s}}[{\rm B}(\omega,z)](\mathcal C(\tau,s)-\chi(s)\mathcal \zeta(s-1))}{\chi(\tau)\chi(s)Q(\tau-1)Q(s-1)}.\label{covmax}
	\end{equation}
	\hrulefill
	\vspace*{4pt}
\end{figure*}
\subsection{Asymptotic consistency with the replica ansatz}
We assume that $\chi(t)$ is convergent as $t\to \infty$. Then, the weak long-term assumption \eqref{weaklong} implies that $\zeta(t)\to 1$. Thereby, it is easy to see from the expressions in \eqref{sigmax} and \eqref{covmax} that $\sigma_x(t)\to 1$ and $\mathcal C_{\theta}(t,t)\to 1/v$. In other words, the field statistics asymptotically obey the equations of the static solutions of the replica ansatz \eqref{replica} \cite{Kab14,Tulino13}.
\subsection{State evolution of AMP}
For  matrix ensemble (i), it follows from  $a_{n}=(\alpha\xi^{-1})\xi^{-n}$ that $\sigma_{ x}(t)=1$ and $\zeta(t)=1$. Moreover, from \eqref{Riid} we get ${\rm B}(\omega,z)=\alpha \omega z\xi^{-2}$. 
Plugging these results into \eqref{covmax} yields the state evolution formula of the AMP algorithm \cite{Donoha,Bayati}:
\begin{equation}
C_{\theta}(t+1,t+1)= \frac{1}{\xi}+ \frac{1}{\alpha}\left\langle \left( \eta_{v(t)}\left(\sqrt{\mathcal C_{\theta}(t,t)}z+ x\right)-x\right)^2\right\rangle\nonumber 
\end{equation}
where $z\sim \mathcal N(0,1)$ is independent of $x\sim p(x)$.
\subsection{Example: Matrix ensemble~(ii)}
For matrix ensemble (ii), we obtain the expression ${\rm  Co}_{\omega^{n}z^{k}}[{\rm B}(\omega,z)]=-(\xi_1\xi_2)^{-n}\delta_{nk}$. Using this result in \eqref{covmax}, we recast the variance $C_{\theta}(t,t)$ in (\ref{covmax}) in the compact form
\begin{equation}
C_{\theta}(t,t)=\frac{\zeta(t)}{v(t)}\sigma_x(t)- \frac{{\rm R}(\chi(t))^2}{\chi (t)^2v(t)^2}\kappa(t)
\end{equation}
with $\kappa(t)=[\chi (t)-\chi(t)\zeta(t-1)+\mathcal{\hat G}(t,t-1)^2\kappa(t-1)]/(\xi_1\xi_2)$. 
\section{Summary and Outlook}
By making use of DFT, we have introduced a theoretical method to design and analyze iterative algorithms for solving the TAP equations of compressed sensing with general invariant matrix ensembles. We have focused our attention on the single-step memory design as it yields the AMP algorithm and its state evolution formula under the classical iid assumption for the entries of the observation matrix. Yet, there are many interesting way of defining other designs, e.g. the single-step response design: $\mathcal G(t,\tau)=0$, for all $\tau\neq t-1$. We leave the study of such schemes to future work. 

We did not include any analysis of the dynamical instability of the algorithm. Our simulation results -- not reported here -- show sensitivity to initialization, especially for high noise precision $\xi$. Augmenting the analysis with an investigation of the dynamical instability conditions will potentially allow us to design more powerful algorithms.

Recently, \cite{rang,Keigo} have reported important results on the dynamical analysis of expectation-propagation based algorithms whose fixed points are consistent with the TAP equations \eqref{TAP}. An interesting task is to relate these results with those obtained with DFT.

\bibliographystyle{IEEEtran}
\bibliography{IEEEabrv,mybib}
\end{document}